\input{psfig}
\documentstyle[preprint,aps]{revtex}
\tightenlines
\begin{document}
\preprint{\begin{minipage}{2in}\begin{flushright}
\end{flushright}
  \end{minipage}}
\title{The $\eta'$ Meson with Staggered Fermions}
\author{L. Venkataraman and G. Kilcup\address{Department of Physics, The Ohio State University, 174 West 18th Ave, Columbus OH 43210}}
\maketitle
\begin{abstract}
We have computed the $\eta'$-pseudoscalar octet mass splitting using 
staggered fermions on both dynamical and quenched gauge configurations.
We have used Wuppertal smeared operators
to reduce excited state contributions. We compare our results
with the theoretical forms predicted by partially
quenched chiral perturbation theory in the lowest order.
Using lattice volumes of size $16^3 \times 32$ with
$a^{-1}=2$GeV we obtain results consistent with the
physical $\eta'$ mass.
We also demonstrate that the flavor singlet piece of
the $\eta'$ mass comes from zero modes of the Dirac operator.

\end{abstract}
\maketitle

\section{INTRODUCTION}
By now there have been dozens of calculations of the masses
of most of the light hadrons using lattice QCD.  However,
until recently the $\eta'$ meson received only scant attention,
in large part because of the relative difficulty of the 
calculation: the disconnected contraction which gives the
$\eta'$ propagator its special character is an order of
magnitude more expensive to compute.  A previous study
\cite{KUR} made a first study of the $\eta'$ two-point
function using quenched Wilson fermions, finding a vertex
of the right size to explain the $\eta'$ mass.  Here we
improve the situation in several ways, by using staggered
fermions which have better chiral properties, by going
closer to the chiral limit, and most importantly by using
gauge configurations with dynamical fermions.
The authors of the earlier study also took a step toward
confirming the conventional wisdom that the $\eta'$
receives its special mass from instantons by sorting
their gauge fields into bins of topological charge.
Here we take the further step of examining the contribution
of topological zero modes of the Dirac operator to the
$\eta'$ propagator.  The results again confirm the lore
that this is the mechanism by which the $\eta'$ gets its mass.

\begin{figure}[htbp]
\centerline{\psfig{figure=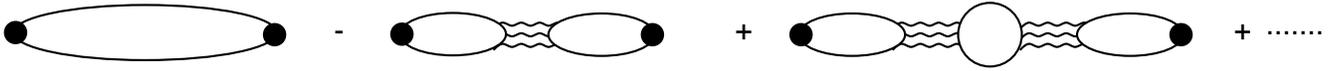}}
\caption{Virtual quark loops in the $\eta'$ propagator}
\label{VQL}
\vspace{0.5cm}
\end{figure}

The $\eta'$ meson is one of the more intriguing strongly interacting
particles.  Although it is the lightest flavor singlet pseudoscalar,
it is too heavy to be the Goldstone boson of a $U(1)$ axial symmetry,
as was pointed out by Weinberg \cite{WEINBOUND} long ago.
With the emergence of QCD it was understood that this $U(1)$
symmetry is actually anomalous, so that one should not
expect an associated Goldstone boson.  
A qualitative understanding of the mechanism behind the
$\eta'$ mass was provided in the framework of the large
$N_c$ expansion.  When $N_c \rightarrow \infty$ ~\cite{WIT,VEN},
$U(1)$ axial symmetry is restored and  gives rise to $\eta'-\pi$
degeneracy.  This degeneracy is lifted by the presence of virtual
quark loops (Fig ~\ref{VQL}) in the $\eta'$ propagator which are 
suppressed by one power of $1/N_c$.
Infinite iteration of these quark antiquark annihilation diagrams
gives rise to a geometric series for the Euclidean $\eta'$ propagator.
Defining the correlator $\langle \eta'(t) \eta'(0) \rangle$
one can write its Fourier transform as
\begin{equation}
G_{\eta'}\,\,\,\,\, \sim\,\,\,\,\, \frac{1}{p^2\,\, + \,\,m_{8}^2}\,\, - \,\,\frac{m_0^2}{p^2\,\, + \,\,m_{8}^2}\,\, + \,\,\frac{1}{p^2 + m_8^2}\,m_{0}^2\,\frac{1}{p^2 + m_8^2}\,m_0^2\frac{1}{p^2 + m_8^2}\,\, + {} \cdots
\label{geom}
\end{equation}
where $m_8^2$ is the average of the square of the octet masses
and $m_0^2$ is the strength of the flavor singlet interaction.
Summing the geometric series shifts the pole from $m_8^2$ to 
$m_0^2 + m_8^2$.  Thus
$$\langle \eta'(t) \eta'(0) \rangle \,\,\,\,\sim \,\,\,\, \frac{1}{p^2\,\, + \,\,m_8^2\,\, + m_0^2} $$
In an $SU(3)$ symmetric world, one would write 
\begin{equation}
m_{\eta'}^2 = m_0^2 + m_8^2
\label{SU3}
\end{equation}
$m_8^2$ vanishes in the chiral limit while $m_0^2$ does not.  Taking
$m_8^2=(4m_K^2+3m_{\pi}^2+m_{\eta}^2)/8$ and substituting the 
experimentally measured values for $m_{\eta'}$, $m_K$, $m_{\pi}$ and
$m_{\eta}$, one obtains $m_0=860{\rm MeV}$.

The large $N_c$ approximation also gives the qualitative understanding
that this $m_0$ is linked to instantons.
To leading order in $1/N_c$ one finds the Witten-Veneziano formula 
\begin{equation}
m_0^2 = \frac{2\,N_f\chi}{f_{\pi}^2}
\label{WV}
\end{equation}
where $N_f$ is the number of flavors, $\chi$ is the topological 
susceptibility and $f_{\pi}$ is the pion decay constant.
In the real world, taking the number of active light flavors to be
three, one needs $\chi$ of order $(185 {\rm MeV})^4$ to explain the
$\eta'$.  Lattice QCD calculations of the quenched topological
susceptibility (e.g. \cite{ALLES}) do indeed give results in
this range.  Thus modulo concerns over $1/N_c$ corrections,
definitions of topological charge, setting the quenched scale
etc., one could claim that lattice calculations give a
quantitatively correct indirect determination of $m_0$.

\section{Direct Lattice Calculation of {$\bfseries m_0$}}

\begin{figure}[htbp]
\centerline{\psfig{figure=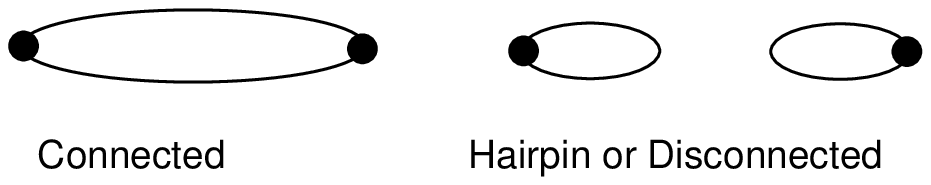}}
\caption{Contributions to $\eta'$ correlator}
\label{corr}
\vspace{0.5cm}
\end{figure}

Lattice QCD offers the challenge of calculating $m_0^2$ directly
from the $\eta'$ propagator without resort to any argument
based on large $N_c$.  The quantity of interest is the
two point function $\langle {\eta'}(x){\eta'(y)}$, where 
the  simplest choice for the interpolating field $\eta'(x)$ is
$\overline{Q}(x){\gamma}_5\,Q(x)$.  As illustrated in figure
~\ref{corr}, there are two types of contractions to take into 
account: (i) the single quark loop connected diagram which 
contributes to flavor-singlet and flavor non-singlet mesons alike,
and (ii) the disconnected diagram which appears only for the
flavor singlet $\eta'$.
As in the previous lattice studies \cite{KUR,KIL,MAS,THA},
we find it convenient to define the ratio $R(t)$ of the disconnected
two loop amplitude to the connected one loop amplitude
\begin{equation}
R(t) = \frac{\langle \eta'(t)\eta'(0){\rangle}_{disc}}{\langle \eta'(t)\eta'(0){\rangle}_{conn}}
\label{RAT}
\end{equation}
Noting the asymptotic behavior
\begin{equation}
{\langle{\eta'(0)\eta'(t)}\rangle
 \over
\langle{\pi(0)\pi(t)}\rangle}
\rightarrow {Z' \exp(-m_{\eta'}t)\over Z \exp(-m_{\pi}t)}
\end{equation}
one sees that in full unquenched QCD the ratio $R(t)$ asymptotes
to $1 - B\exp(-\Delta m t)$,
where B is a constant and $\Delta m = m_{\eta'} - m_8$.
This statement assumes that there are equal numbers
of dynamical and valence flavors and $m_{dyn}=m_{val}$.
If one has differing numbers of valence and dynamical fermions,
as we do when using staggered fermions, then one needs to
rescale the connected diagram
of fig~\ref{corr} by $N_f/N_v$, and the disconnected diagram
by $(N_f/N_v)^2$.  Therefore $R(t)$ takes the form
\begin{equation}
R(t) = \frac{N_v}{N_f}\,\,\lbrack 1\,\, - \,\,B \exp(-\Delta mt)\rbrack
\label{DRAT}
\end{equation}

On quenched configurations, the absence of closed quark loops
means that the basic vertex is not allowed to iterate, and
equation~\ref{geom} is truncated after the first two terms.
Thus in the quenched  approximation, there is a double pole
in the $\eta'$ propagator.
This means for the zero spatial momentum state, $R(t)$ rises
linearly with the slope $\frac{m_0^2}{2m_8}\,t$.
\begin{equation}
R(t) = const. + \frac{m_0^2}{2m_8}\,t
\label{QRAT}
\end{equation}

We have used staggered fermions, both dynamical and quenched
configurations and local and Wuppertal smeared operators
for the extraction of $m_{\eta'}$. 
We also study the dynamical flavor dependence of 
$m_{\eta'}$  and have derived the expected theoretical forms
for the ratio using partially quenched chiral perturbation theory
($PQ\chi PT$) so as to enable comparison of our results with theory.
In section 3, we review the basic concepts of $PQ\chi PT$~\cite{PQPT}
and derive expressions for $R(t)$. In section 4, we describe 
details pertaining to the parameters of the simulation and the
Wuppertal smearing procedure.
In section 5 we discuss the results obtained from our simulation
and compare with the theoretical predictions of section 2.

\section{RATIO from $PQ\chi PT$}

Bernard and Golterman~\cite{QPT,PQPT} have developed a technique
for constructing an effective chiral theory for quenched and
partially quenched QCD. The basic idea in this approach is based
on the observation that if a scalar quark ($\tilde{q}$) is added to 
the QCD Lagrangian, then the scalar determinant can cancel
the quark determinant. For concreteness, if we assume 
there are $N_v$ quarks of masses $m_i(i=1,2..N_v)$, then adding 
$k$ pseudoquarks of masses $m_j(j=1,2..k)$ such that $m_j=m_i$
would mean that first $k$ quarks are quenched and the remaining
$N_v-k$ unquenched.
The low energy effective chiral theory then
describes interactions among all kinds of 
boundstates, including the ordinary pions as well
as unphysical states containing scalar quarks.
The reader is referred
to ~\cite{QPT,PQPT} for discussion regarding the form of the
Lagrangian and its symmetry properties. For the purposes of
this paper, it is sufficient to write down the Green's
function in momentum space in the basis of the states corresponding
to $\overline{u}u$, $\overline{d}d$  and their pseudoquark counterparts:
\begin{equation}
G_{ij} = \frac{{\delta}_{ij}{\epsilon}_i}{p^2 + M_i^2} - \frac{m_0^2}{(p^2
 + M_i^2)(p^2 + M_j^2)F(p^2)}
\label{GF}
\end{equation}
where
\begin{equation}
F(p^2) = 1 + m_0^2\sum_{d=k+1}^{N_v}\frac{1}{p^2 + M_d^2},\nonumber
\end{equation}
$M_i^2$ is the square of the meson mass composed of quark flavor $i$ and
\begin{equation}
{\epsilon}_i = \cases{  +1 & {$1 \leq i \leq N_v$} \cr
-1 & {$N_v+1 \leq i \leq N_v+k$}}\nonumber
\end{equation}

In the above, the term proportional to $\alpha$, the 
momentum dependent self-interaction of $\eta'$ has
been neglected. It can be reinstated any time by the substitution
$m_0^2 \rightarrow m_0^2 + {\alpha}^2 p^2$.
When $m_i=m_j$ the first term is simply the neutral meson 
propagator in the absence of flavor singlet interactions.
The second term is obtained from summing the infinite geometric series
obtained by iterating the flavor singlet interaction.  When simulating
with staggered fermions on $N_f=2$ dynamical configurations, the
variables $N_v$ and $k$ take the values 4 and 2 respectively.
Specifically, when all the valence flavors have the same mass $m_{val}$
and all the dynamical flavors have mass $m_{dyn}$ but $m_{val} \ne m_{dyn}$,
(in a typical simulation this situation would arise if one
were simulating with staggered fermions on dynamical configurations)
the Green's function above takes the form,
\begin{equation}
G_{ii} = \frac{1}{p^2 + M_i^2} - \frac{m_0^2\,\,\,(p^2 + M_d^2)}{(p^2 + M_i^2)^2\,\,(p^2 + M_d^2 + N_fm_0^2)}
\label{releq}
\end{equation}
It is straightforward
to go over to configuration space, project each term on to
zero spatial momentum state and then take the ratio of the second
term to the first term\footnote{All the formulae have been derived 
in the infinite volume limit}.  Doing so we obtain

\begin{equation}
\label{PQRAT}
R(t)  = \,\,\frac{N_v}{N_f}\,\lbrack\, At\,\, -\,\, B\exp(-\Delta mt)\,\, + \,\,C\rbrack
\end{equation}
where 
\begin{equation}
\hspace{-250pt}
A = \frac{(m_d^2 - m_8^2)(m_d^2 - m_{\eta'}^2)}{(m_8^2 - m_{\eta'}^2)2m_8},
\end{equation}
\begin{equation}
\hspace{-280pt}
B = \frac{m_8(m_d^2 - m_{\eta'}^2)^2}{m_{\eta'}(m_8^2 - m_{\eta'}^2)^2},\\
\end{equation}
\begin{equation}
\hspace{-135pt}
C = \frac{(m_d^2 - m_{\eta'}^2)}{(m_8^2 - m_{\eta'}^2)}\,\left[\frac{(m_d^2 - m_8^2)}{2m_8^2}\,+\, \frac{(m_d^2 - m_8^2)}{(m_8^2 - m_{\eta'}^2)} + 1\right],
\end{equation}
\begin{equation}
\hspace{-226pt}
m_8^2\equiv M_i^2=M_j^2,\,\,\, m_d^2 \equiv M_d^2\,\,\,\, \rm{and}
\end{equation}
\vspace*{-37pt}
\begin{equation}
\hspace{-294pt}
m_{\eta'}^2 = m_d^2 + N_f\,m_0^2.
\end{equation}

As a check we note that eqn~\ref{PQRAT} contains 
the expression in eqn~\ref{DRAT} and eqn~\ref{QRAT} as 
appropriate limits.
When $m_8=m_d$, one obtains eqn~\ref{DRAT} with 
$B=m_8/m_{\eta'}$,
and in the limit 
$m_d \rightarrow \infty$, $N_f \rightarrow 0$ one
obtains eqn~\ref{QRAT} with the intercept
equal to $\frac{m_0^2}{2m_8^2}$.   
We have ratio data on quenched and $N_f=2$ configurations.
Our ratio data on $N_f=2$ configurations can be further divided
into $m_{val}=m_{dyn}$ and $m_{val} \ne m_{dyn}$.  Quenched
and $m_{val} = m_{dyn}$ data can be subjected to both one and
two parameter fits thus enabling comparison between the predictions 
of the theory and ``experiment''.
Likewise, a similar comparison is possible from both
one and four parameter fit to $m_{val} \ne m_{dyn}$ data.  

\section{SIMULATION DETAILS}

\subsection{Ensemble}

The propagators required for computing the disconnected and
the connected amplitude were computed on configurations of
size $16^3 \times 32$. The statistical ensemble used
and the valence quark masses at which the simulation was performed
are shown in table~\ref{TAB1}. The dynamical configurations
were borrowed from Columbia while the quenched configurations
were generated on the Cray T3D at Ohio Supercomputer Center(OSC).
By design, the ensembles listed in table~\ref{TAB1}
have the same inverse lattice spacing of about $2{\rm GeV}$ obtained
from $m_{\rho}$. The values of $m_{val}$ on the quenched configurations 
were chosen 10\% higher than those on the dynamical configurations,
in order to bring the corresponding values of $m_8$ into more precise
agreement.
The staggered propagators were computed by invoking  the 
conjugate gradient method 
using sets of processors ranging from 16 to 64 nodes 
of the Cray T3D at OSC and Los Alamos's Advanced Computer Laboratory
with a sustained performance of 45 Mflops per node.

\subsection{Propagators}

The pseudoscalar operator that creates or destroys an $\eta'$ 
in the staggered formalism is $\overline{Q}{\gamma}_5\otimes IQ$.
It is a distance 4 operator in which the quark and the antiquark 
sit at the opposite corners of the hypercube. We put in explicit links
connecting the quark and the antiquark across the edges of the hypercube
and averaged over all the 24 possible paths to make 
the operator gauge invariant.
As was mentioned earlier, the two point correlation function of this 
operator yields both the connected one loop and the disconnected
two loop amplitude of fig~\ref{corr}. Two propagators need to be computed
for calculating the one loop amplitude. One propagator was 
calculated using a delta function source
with a local random phase ${\eta}_x=\exp(i\theta (x))$ at each
spatial site and for each color on time slices $t=1$ and $t=2$. 
When averaged over all noisy samples, the noise ${\eta}_x$ satisfies
$\langle {\eta}_x{\eta}_y^{\dag}\rangle = {\delta}_{xy}$.     
Translating the noisy source at each site (made gauge invariant by putting
in links) a hypercubic distance 4 and putting in the phase appropriate
for the $\eta'$ operator, the second propagator was calculated. 
Transporting the source and calculating the anti-propagator
separately is the price paid due to the non locality of the 
staggered flavor singlet operator. For each configuration we took
two noise samples with the lattice doubled in the time direction.

Very few lattice calculations of $m_{\eta'}$ exist because of the
difficulty involved in getting good signals for the disconnected amplitude
using reasonable computer time.
We have addressed this problem
by using a noisy source like the one used for the connected amplitude
but placed on all sites of the lattice.  Then we solve 
$(\rlap{\,/}D + m)\phi = \eta$ and  estimate the
quark propagator as $G_{xy} = \langle m {\phi}_x{\phi}_y^{\dag}\rangle$
with $\vert x-y\vert$ even. The distance 4 staggered flavor singlet 
operator that we use satisfies this criterion. 
At this point it should be mentioned that there also exists 
a distance 3 staggered flavor singlet operator for which the
the appropriate estimator of 
$G_{xy}$ is $\langle {\eta}_x^{\dag}{\phi}_y\rangle$.  
This channel did not yield good signals for the 
the pseudofermion propagators, and we did chose not to
examine it further.
In any case, $\langle m {\phi}_x{\phi}_y^{\dag}\rangle$ is a better
estimator than $\langle {\eta}_x^{\dag}{\phi}_y\rangle$ since the
fluctuations in the former go as ${\langle\overline{\chi}\chi\rangle}^2$
while those for the latter are of the order 
$\langle \overline{\chi}\chi\rangle/m_{val}$ where 
${\langle\overline{\chi}\chi\rangle}$ is the condensate. For the
range of light quark masses chosen in this simulation, 
${\langle\overline{\chi}\chi\rangle}$ lies between 0 and 1, thus
making 
${\langle \overline{\chi} \chi\rangle}^2$ significantly less than
$\frac{\langle \overline{\chi} \chi\rangle}{m_{val}}$.
We used 16 noise samples per configuration, again on a doubled
lattice.

\subsection{Wuppertal Smearing}

\begin{figure}[hbtp]
\centerline{\psfig{figure=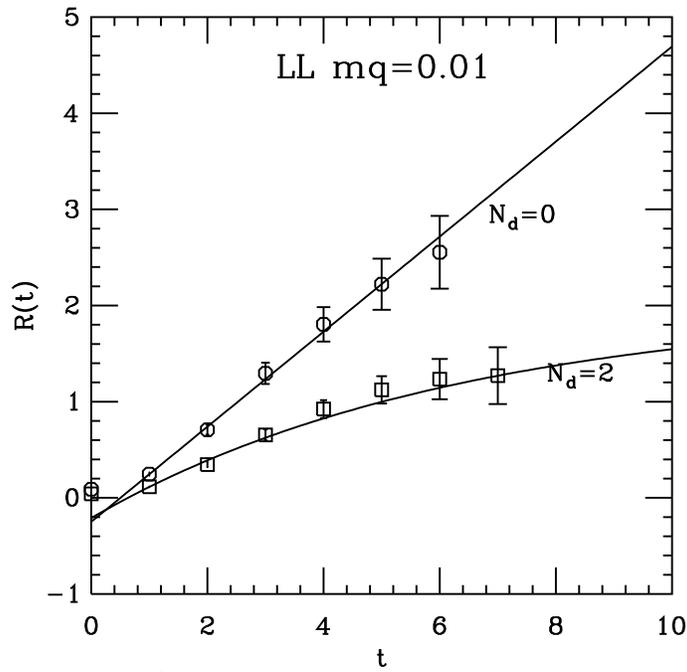,width=9.0cm}}
\vspace{-0.2cm}
\caption{Ratio using local operators.}
\label{tyRAT}
\end{figure}

Fig~\ref{tyRAT} shows the ratio $R(t)$ on quenched and
dynamical configuration ($N_f=2, m_{dyn}=0.01$) using
local interpolating fields.
Since the disconnected data is noisy, the ratio data is 
only useful for the first few time slices where the contribution
due to excited states is non-negligible.  Smeared operators
reduce the contributions from excited states and enable a more
reliable extraction of ${\Delta}m$ from the available data.

We used the Wuppertal smearing technique~\cite{SOM} in which 
one obtains an exponentially decaying bound state wavefunction
from solving 
\begin{equation}
(-{\nabla}^2[U] + {\kappa}^2)\phi(x) = {\delta}_{x,0}
\label{WUP}
\end{equation}
where ${\kappa}^2$ is a parameter that can be tuned to control
the spread of the wavefunction. On the lattice, for staggered
fermions and to make the smearing procedure gauge invariant, the
operator ${\nabla}^2$ takes the form
\begin{eqnarray}
{\nabla}^2[U] & \rightarrow & \frac{1}{2}\sum_{\mu =1}^{3} U_{\mu}(x)\,\,U_{\mu}(x+\mu)\,\delta_{x',x+2\mu} +\,\,U_{\mu}^{\dagger}(x-\mu)\,U_{\mu}^{\dagger}(x-2\mu)\delta_{x',x-2\mu}\,\, - \,\,2{\delta}_{x',x}
\label{LWUP}
\end{eqnarray}

\begin{figure}[hbtp]
\centerline{\psfig{figure=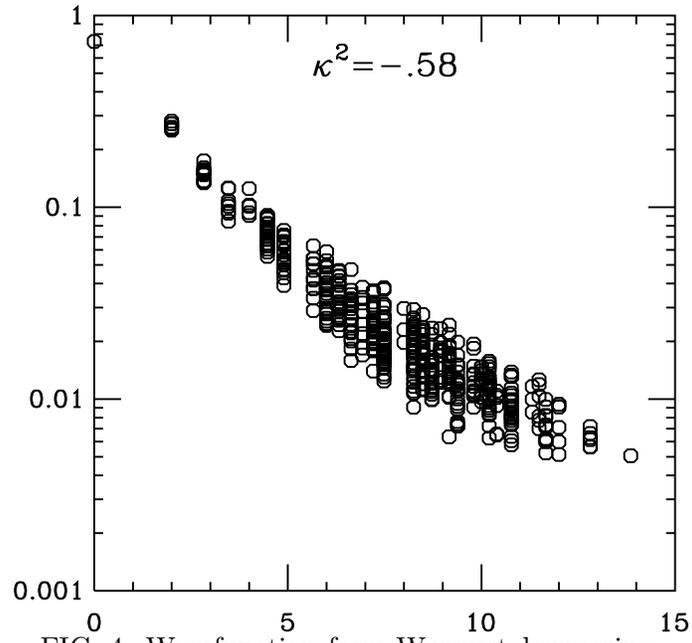,width=9.0cm}}
\vspace{-0.2cm}
\caption{Wavefunction from Wuppertal smearing.}
\label{WF}
\end{figure}

We experimented with different values of ${\kappa}^2$ and determined
that the  critical value occurred near ${\kappa}^2=-0.64$. The values
of ${\kappa}^2$ that are relevant to this study are those near
the critical value since this corresponds to a maximum spread
in the wavefunction without losing the exponential behavior.
The form of the wavefunction obtained on one of the $N_f=2(m_{dyn}=0.01)$ 
configurations is shown in fig~\ref{WF}. 

\begin{figure}[hbtp]
\centerline{\psfig{figure=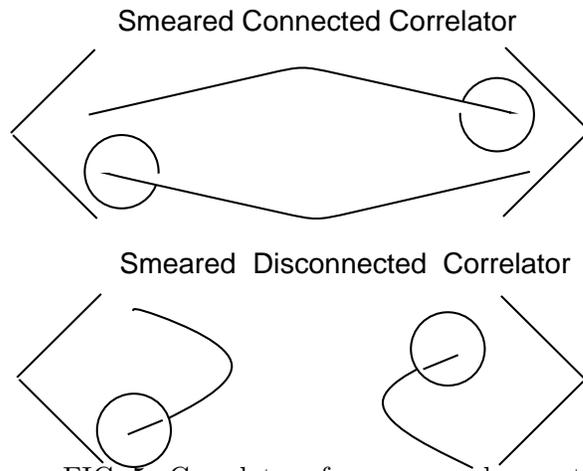,width=9.0cm}}
\vspace{-0.3cm}
\caption{Correlators from smeared operators.}
\label{SCORR}
\end{figure}

For this study, one propagator of the connected correlator 
was calculated with the source smeared to a fixed radius corresponding
to ${\kappa}^2=-0.6$ and a point-like sink.  This was tied to five
different anti-propagators with point-like source  but
smeared at the sink end corresponding to different values of ${\kappa}^2$
(see fig~\ref{SCORR}).  
Thus the  smeared valence propagators that we have calculated 
correspond to the set of correlators shown in table~\ref{TAB2}.

\begin{figure}[hbtp]
\centerline{\psfig{figure=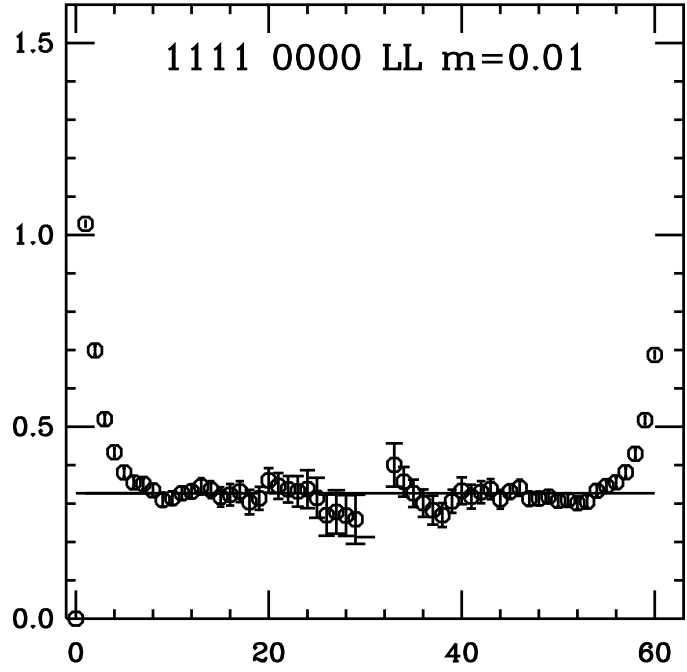,width=9.0cm}}
\caption{Effective mass plot for $m_8$ with LL correlator.}
\label{emLL}
\end{figure}

\begin{figure}[hbtp]
\leftline{\psfig{figure=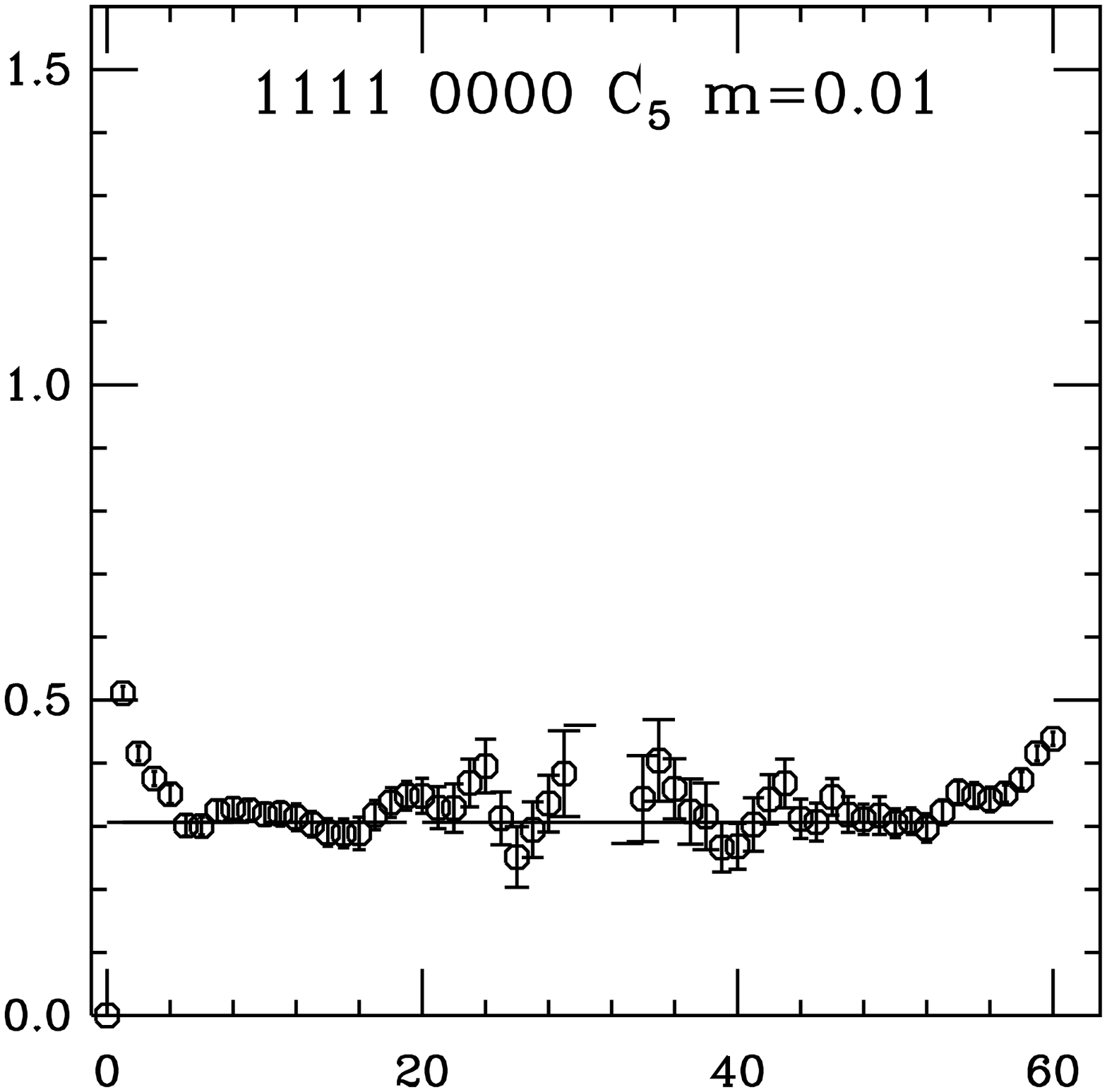,width=6.0cm}
\hskip 3.0cm
{\psfig{figure=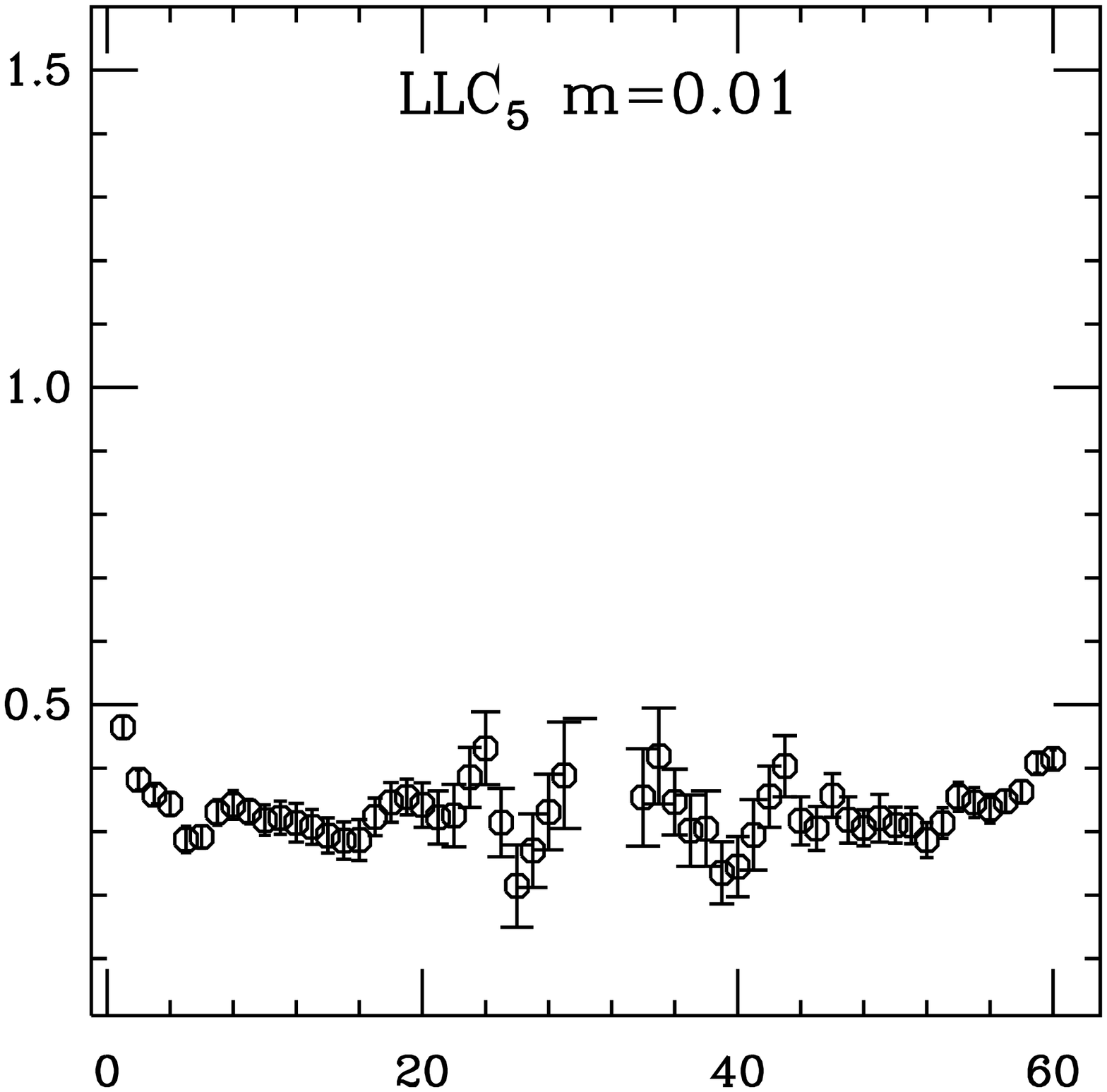,width=6.0cm}}}
\caption{Effective mass plot with $C_5$ and $LLC_5$ correlators.}
\label{emO5}
\end{figure}

A typical effective mass plot
obtained on $N_f=2$, $m_{dyn}=0.01$ configuration at $m_{val}=0.01$
is shown in fig~\ref{emLL} for the correlator LL.  
Since decrease in ${\kappa}^2$ increases the
spread in the wavefunction, among the 5 correlators shown in 
table~\ref{TAB2}, $C_5$ is expected to perform the best which
is evident by comparing fig~\ref{emLL} and fig~\ref{emO5}.
We also defined the correlator 
\begin{equation}
\langle \cos\theta \,\overline{Q}(x)Q(x) + \sin\theta \,\overline{Q}_{w4}(x)Q(x) \rangle 
\label{Lcop}
\end{equation}
which amounts to taking the linear combination 
\begin{equation}
({\cos}^2\theta\, LL+{\sin}^2\theta \,C_5 +2\cos\theta \,\sin\theta C_1)
\label{Lcorr}
\end{equation}
We determined the angle $\theta$ for which the value of the 
effective mass on $t=1$ was the lowest and this corresponded to
$\theta=.797\pi$. We denote this linear combination of correlators
as $LLC_5$ and it is clear from fig~\ref{emO5} that it
is slightly better than $C_5$.

To calculate the corresponding disconnected contributions of the 
correlators listed in ~\ref{TAB2}, the sink end
of the pseudofermion propagator ($\phi'$) was smeared to 4 different 
radii corresponding to 
${\kappa}^2 = -0.5,\,\, -0.53, \,\, -0.56$ and $-0.6$  (see fig~\ref{SCORR}).
Accordingly, the quark propagator was obtained from the estimator
$G_{xy}' = \langle m {\phi}_x'{\phi}_y^{\dag}\rangle$.

\begin{figure}[hbtp]
\centerline{\psfig{figure=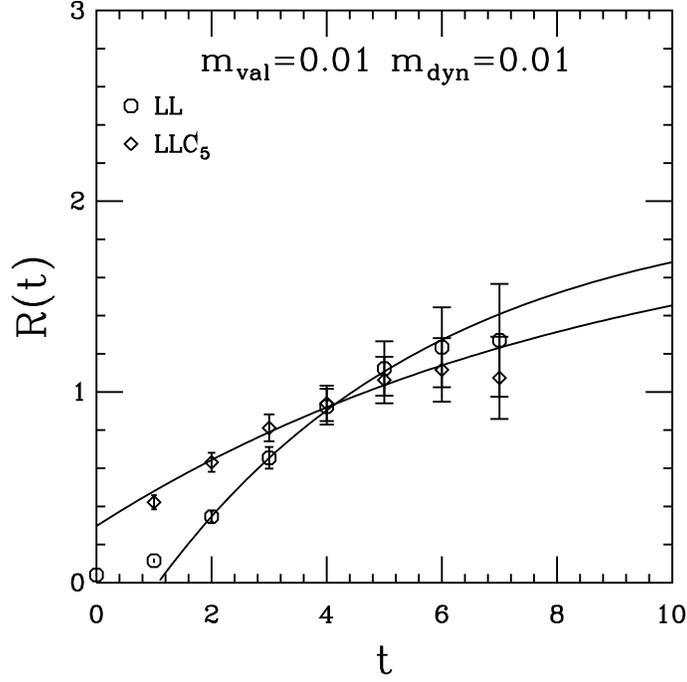,width=9.0cm}}
\caption{Dynamical ratio from local and smeared operators.}
\label{LS}
\end{figure}

Fig~\ref{LS} compares the ratio plot with and without smearing
on dynamical configurations. In the initial time slices both
the data are different but they begin to coincide after a few
time slices as they should.

\section{RESULTS}

\begin{figure}[hbtp]
\centerline{\psfig{figure=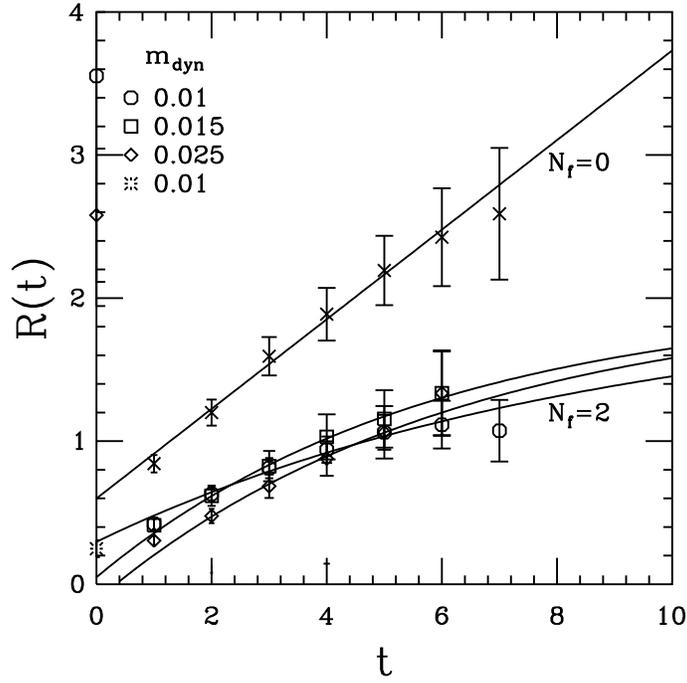,width=9.0cm}}
\caption{Flavor dependence of $R(t)$.}
\label{Nfdep}
\end{figure}

In the plot of figure~\ref{Nfdep}, the different curves 
represent the fit to the ratio data obtained from the various
configurations listed in table~\ref{TAB1}.
For $N_f=2$  the points shown correspond to $m_{val}=m_{dyn}$ 
while for the quenched data the points correspond to $m_{val}=0.011$.
The quenched data is fit linearly according to equation~\ref{QRAT}
with both the intercept and the slope as free parameters.
The dynamical data is fit according to equation~\ref{DRAT}
with both $m_{\eta'}$ and $B$ as free parameters.  A remarkable
observation that is to be gleaned from this graph is that 
the ratio data is distinctly different for different numbers 
of dynamical flavors and in accordance with the predicted 
theoretical form.  If nothing else, this observable is clearly
able to distinguish quenched from dynamical configurations.

\begin{figure}[hbtp]
\leftline{\psfig{figure=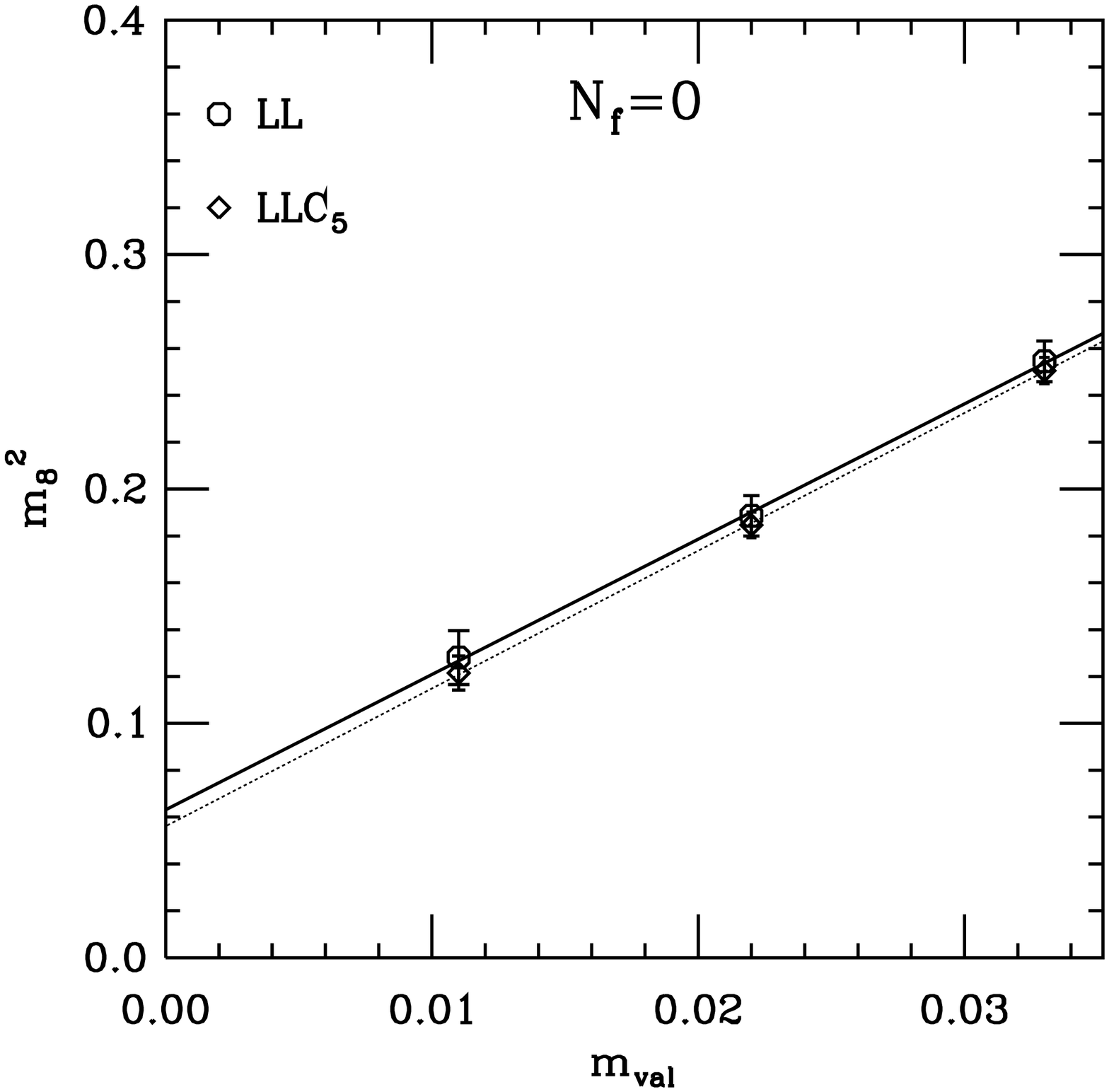,width=6.0cm}
\hskip 2.0cm
{\psfig{figure=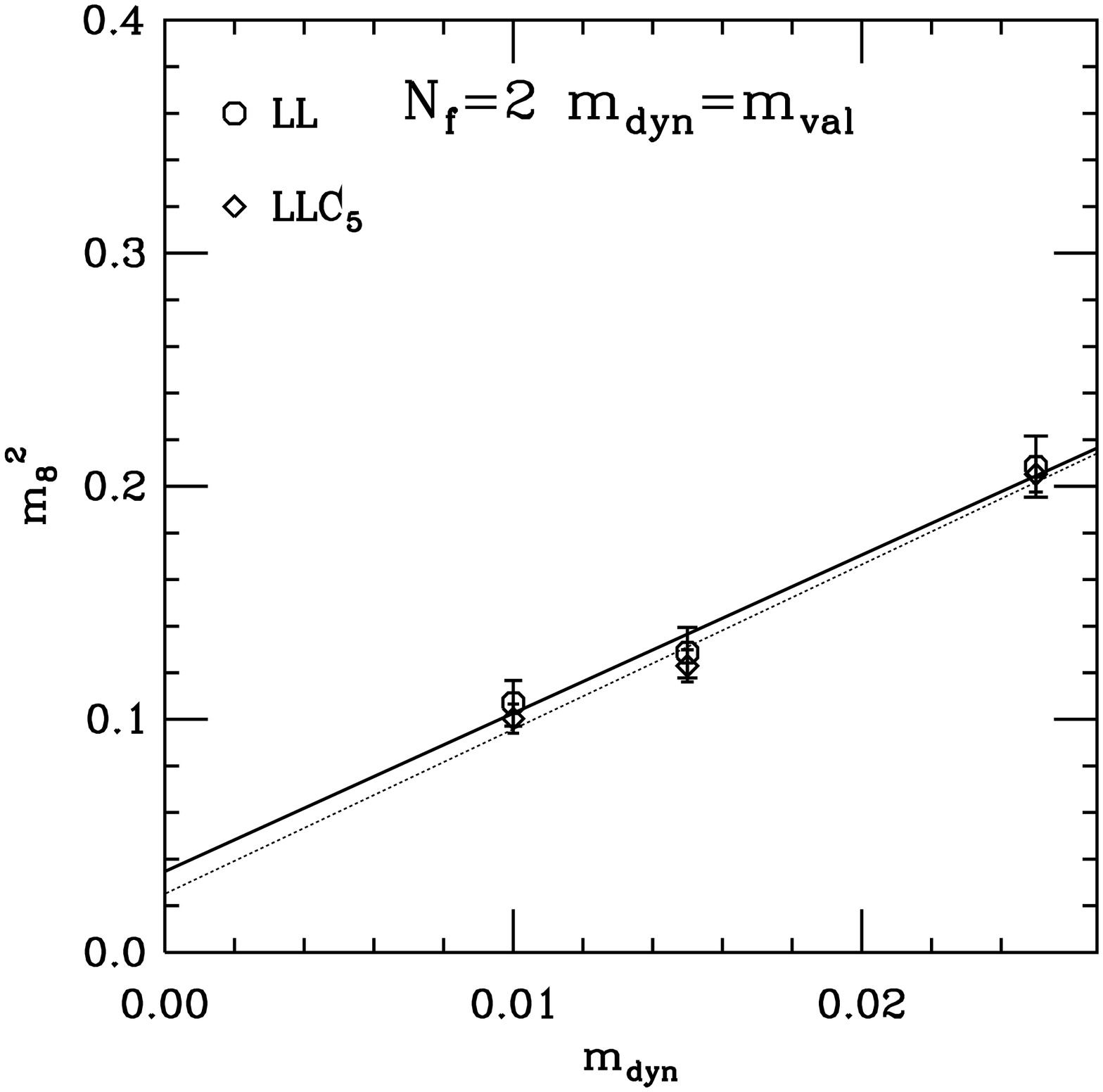,width=6.0cm}}}
\caption{$m_8^2$ vs $m_{val}$ on $N_f=0$ and $N_f=2$ configurations.}
\label{Dm8}
\end{figure}

Our values of $m_8$ calculated  on quenched and $N_f=2$ dynamical 
configurations are plotted as a function of the valence quark mass
in fig~\ref{Dm8}.  We note that at this finite lattice spacing,
$m_8$ does not quite vanish in the chiral limit.  This is a
consequence of our taking the flavor singlet ($\gamma_5\otimes I$)
pion as opposed to the special (flavor $T_5$) staggered pion.
Of course even for our pion, the intercept should vanish in the 
continuum limit.

\begin{figure}[hbtp]
\centerline{\psfig{figure=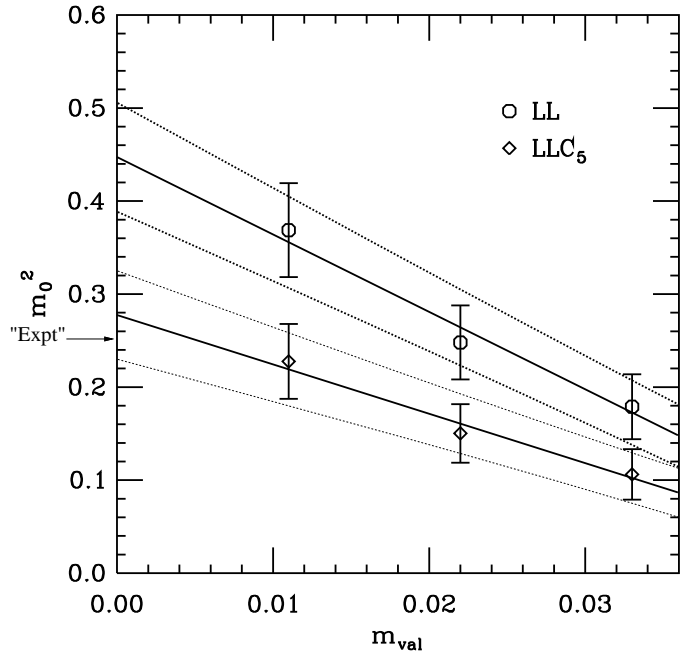,width=9.0cm}}
\caption{Chiral extrapolation of quenched $m_0^2$.}
\label{m02}
\end{figure}

From the linear fit to the quenched data for $m_{val}=0.011,\,\,\,
0.022\,\,\, {\rm and}\,\,\, 0.033$, we extract $m_0^2$ from the slope.
A plot of $m_0^2$ against $m_{val}$ and the extrapolation to the zero
quark mass is shown in fig~\ref{m02}.  $m_0^2$ increases with
decreasing quark mass and does not vanish in the chiral limit.
Smearing has lowered the value of the intercept obtained in
the zero quark mass limit, bringing it closer to the real world value.

\begin{figure}[hbtp]
\centerline{\psfig{figure=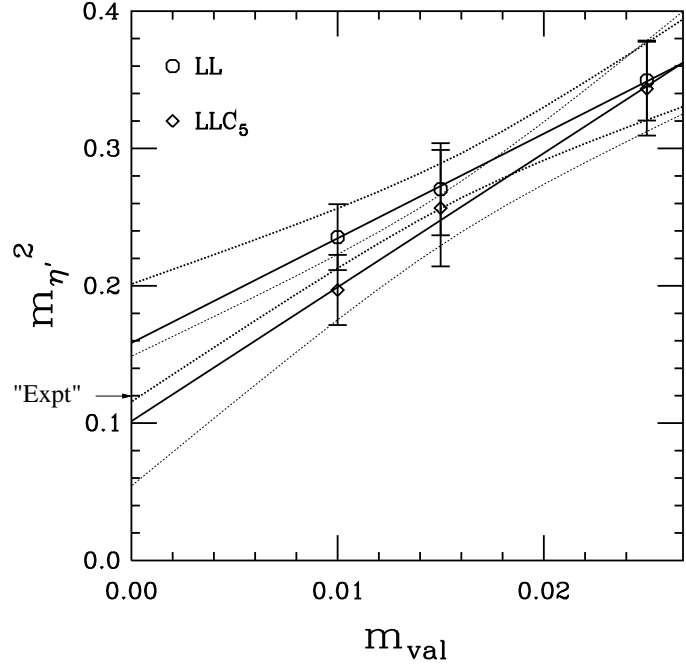,width=9.0cm}}
\caption{Chiral extrapolation of ${m_{\eta'}}^2$ from $N_f=2$ configurations.}
\label{meta}
\end{figure}

Two parameter fits, according to eqn~\ref{DRAT}, of the
dynamical $m_{val}=m_{dyn}$ data yields $\Delta m$ and hence
$m_{\eta'}$. Extrapolating the values of $m_{\eta'}$ so extracted,
we find that it does not vanish in the chiral limit (fig~\ref{meta}).
As seen in quenched case, the values of $m_{\eta'}$ both at finite
and zero quark mass due to the correlator $LLC_5$  are lower than 
the corresponding points due to the local correlator LL.
Errors quoted on the ratio data and all the mass values are obtained
by doing a single elimination jackknife.

\begin{figure}[hbtp]
\centerline{\psfig{figure=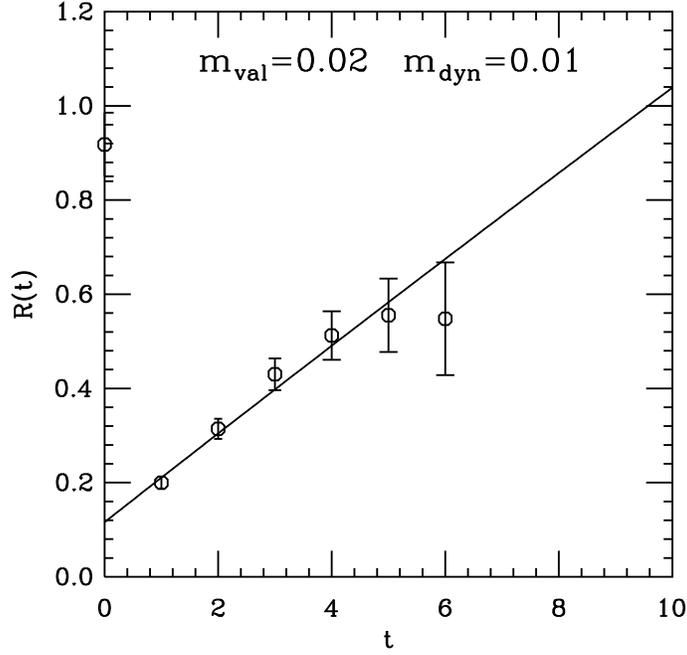,width=9.0cm}}
\caption{One parameter fit to $R(t)$ with $m_{val}=0.02$ and $m_{dyn}=0.01$.}
\label{R0.02}
\end{figure}

We could not obtain stable 4 parameter fits to our $m_{val} \ne m_{dyn}$
data.  The known theoretical forms for the ratio obtained from 
lowest order $PQ\chi PT$ were used for one parameter fit of 
the $m_{val} \ne m_{dyn}$ and $m_{val} = m_{dyn}$ data. 
Fig~\ref{R0.02} shows the theoretical fit obtained for $m_{val} =0.02$
on $N_f=2$ , $m_{dyn}=0.01$ configuration using smeared operators.
One parameter fit to the data obtained with local operators
gave reasonable ${\chi}^2$ only if the fit range began from $t \ge 3$.
This suggests that the local data is heavily contaminated with excited
states since the formulae derived in section 2 hold for asymptotic times.
Doing a similar job
for $m_{val}=0.03$ data and extrapolating the value of $m_{\eta'}$
obtained from both to the chiral limit we obtain 
$m_{\eta'}(N_f=3)=876 \pm 16 {\rm MeV}$ which is remarkably
consistent with the value obtained from fully quenched and
dynamical data (table~\ref{TAB3}). 

\begin{figure}[hbtp]
\centerline{\psfig{figure=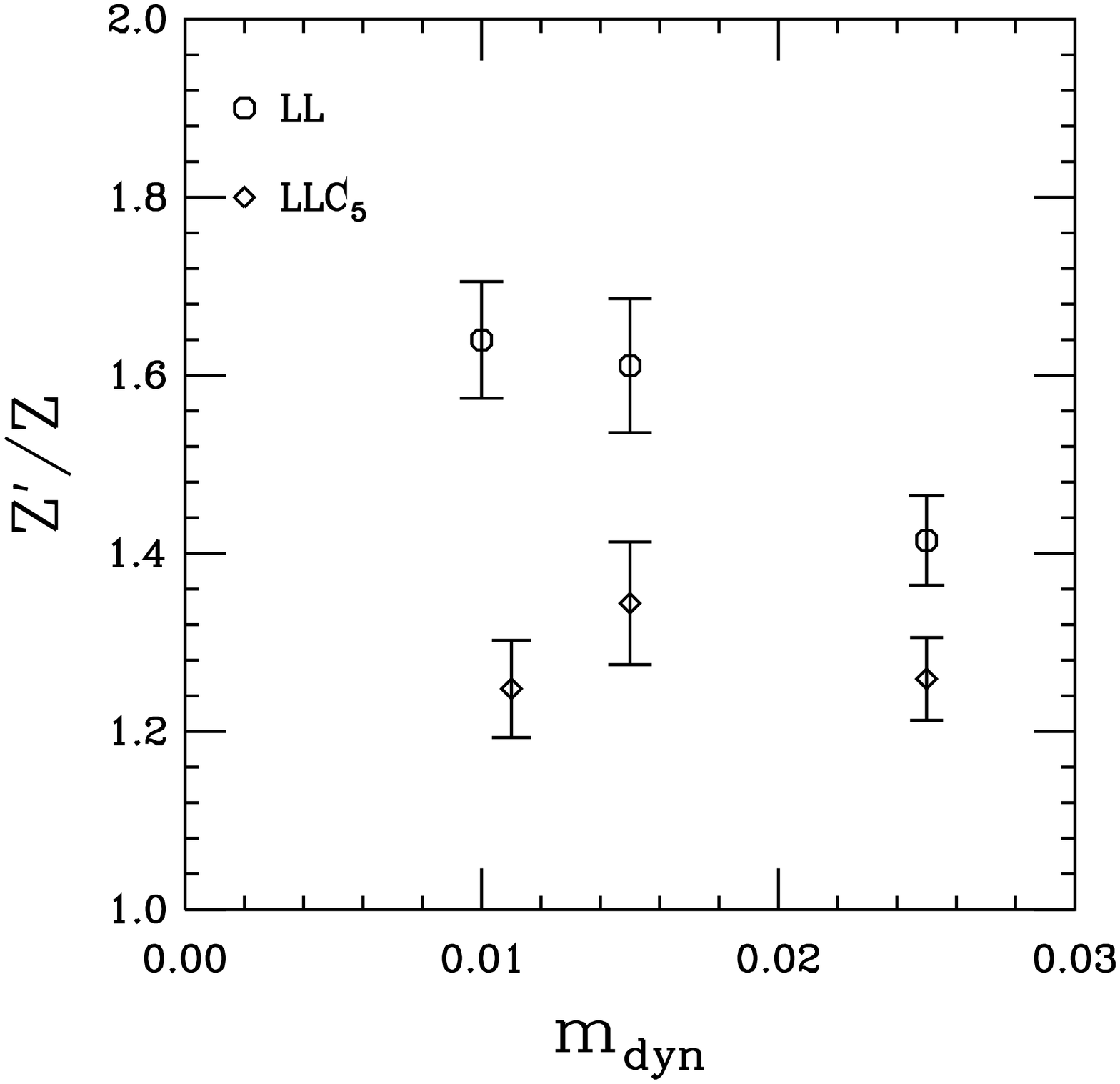,width=9.0cm}}
\caption{$Z'/Z$ vs. $m_{val}$.}
\label{Z'/Z}
\end{figure}

A one parameter fit to $m_{val} \le m_{dyn}$ data did not yield
reasonable ${\chi}^2$.  However, for the $m_{val}=m_{dyn}$ data,
it was possible to extract the deviation from the lowest order
$PQ\chi PT$ as determined by the observable $Z'/Z$, the ratio of
the residues for creating ${\eta'}$ and ${\eta}$.  
\begin{equation}
\frac{Z'}{Z} = \frac{{\mid\langle 0\mid\eta'(0)\mid\eta'\rangle\mid}^2}{{\mid\langle 0\mid\eta(0)\mid\eta\rangle\mid}^2}
\label{Z}
\end{equation}
In $PQ\chi Pt$, this ratio is 1 based on the assumption $f_{\eta'} = f_{\eta}$.
Therefore, in $PQ\chi PT$ as shown in section 2, the parameter B in 
eqn~\ref{DRAT} equals $\frac{m_8}{m_{\eta'}}$ instead of 
$\frac{m_8 Z'}{m_{\eta'} Z}$.  From the 2 parameter fit to the
$m_{val}=m_{dyn}$ data one can then determine the ratio $Z'/Z$. 
The values we extract for $Z'/Z$ are plotted as a function of $m_{val}$
in fig~\ref{Z'/Z} both for smeared and local data.  
It can be seen that the data prefer to be 20-30\% above unity which
is typically the size of higher order chiral and $O(1/N_c)$ corrections.

In the analysis so far, we have neglected the effect of momentum
dependence of $m_0^2$, ie. we had been working in a theory
with $\alpha$, the coefficient of the kinetic energy term of $\eta'$
in the chiral Lagrangian~\cite{PQPT}, set to zero. 
$\alpha$ is expected to be a small quantity, contributing only
at the next to leading order in a combined expansion in $1/N_c$
 and quark mass~\cite{QPT}.
Including $\alpha$ simply shifts the strength of the flavor singlet
interaction vertex from 
$m_0^2$ to $m_0^2 + {\alpha}^2 m_8^2$.   While the form of $R(t)$
remains the same as before for all the three cases, the coefficients
of the time dependent terms and the constants become functions
of two unknown parameters, $m_0^2$ and $\alpha$. 
For the quenched case, we obtain
\begin{equation}
R(t) = \frac{(m_0^2 - \alpha\,m_8^2)\,t}{2\,m_8}\,\,+\,\,\frac{(m_0^2 + \alpha\,m_8^2)}{2\,m_8^2}
\label{Q2prm}
\end{equation}
It is clear from the above formula that the value of $m_0^2$
is shifted by a small amount at finite quark mass and indeed
this is borne out true by the quenched ratio data when fit to
the above form. We see 10\%-15\% downward shift in the 
the values of $m_0^2$ obtained from the slope and 
the intercept of the fit at non zero quark mass. When extrapolated to
the chiral limit, the value of $m_0^2$ is left unaffected (within 
quoted errors in table~\ref{TAB3}) by the presence of the momentum
dependent interaction term. The statistical errors associated
with the extracted values of $\alpha$ are rather large. Our
results for $\alpha$ for each quark mass is shown in table ~\ref{TAB4}.

On dynamical configurations, the presence of a non zero $\alpha$
is reflected in the expression for $m_{\eta'}$. It now takes the form
\begin{equation}
m_{\eta'}^2 = \frac{N_f\,m_0^2 + m_8^2}{1 + N_f\,\alpha} 
\label{meta2p}
\end{equation}
The natural quantity that is extracted from fitting the dynamical
$m_{val} = m_{dyn}$ data with equation~\ref{DRAT} is $m_{\eta'}$
which cannot be used to determine both $m_0^2$ and $\alpha$ 
independently. 

\section{FERMIONIC ZEROMODES}
As reassuring as it may be that our lattice results correctly
reproduce the physical $\eta'$ mass, the calculation itself does not
illuminate the precise mechanism by which the $\eta'$ gets its special
mass.  The conventional lore, as quantified by the Witten-Veneziano
formula, associates $m_0$ with the fluctuations in the topological
charge. In a previous study, Kuramashi et al. tried to verify this
relation by calculating the topological susceptiblity of their gauge
configurations and using it to calculate $m_0$.  They obtain a value
that is higher than the experimantal number.  Here we chose to
illuminate the question by focusing directly on the fermionic
zero-modes which are associated with the topological charge
of  a gauge configuration via the index theorem.

In the continuum, integrating the anomaly equation gives
the relation
\begin{equation}
Q = m\,{\rm tr}\,({\gamma}_5\, S_F)
\label{qfrmu1}
\end{equation}
where $S_F$ is the fermion propagator.
Resolving the propagator in a sum over eigenmodes of $\rlap{\,/}D$,
and noting that all modes with nonzero eigenvalue come in
conjugate pairs, one gets the index theorem:
\begin{equation}
Q = n_+ - n_-.
\end{equation}
Here $n_+$ ($n_-$) is the number of positive (negative)
chirality zeromodes.  If the $\eta'$ is particularly
sensitive to the topological charge of a configuration,
then one must also expect it to be sensitive to the
presence or absence of fermionic zeromodes.

On the lattice, both sides of the index equation are slightly distorted.
The flavor singlet $\gamma_5$ does not exactly anticommute
with $\rlap{\,/}D$, so the trace of $\gamma_5 S_F$ cannot
be collapsed into a trace in the zeromode sector alone.
Further, there are no exact zeromodes, nor is there an exact
definition of the topological charge.  Nevertheless, if we
are sufficiently close to the continuum, something like
the conventional picture should obtain.

We have taken a look~\cite{LAT97} by constructing the lowest few eigenmodes
of $\rlap{\,/}D$ on both the quenched and dynamical
($N_f=2$, $m_{dyn}=.01$) ensembles of gauge fields.
We use the method of subspace iterations as described in
ref. \cite{SUBSPACE}.  This algorithm obtains the eigenmodes
of $\rlap{\,/}D^2$ by minimization of the Ritz functional
\begin{equation}
\mu(X_e) = {\mathop{\rm Tr}} \left[ (X^{\dag}_e X_e)^{-1}  (X^{\dag}_e
(-{\rlap{\,/}D}^2) X_e)\right].
\end{equation}
Here $X_e$ is the rectangular matrix of $m$ eigenvectors.
The subscript indicates we only need the solution on the
even sites; full eigenmodes of ${\rlap{\,/}D}$ can then 
be easily reconstructed on the odd sites.
We found it necessary to modify the algorithm slightly,
periodically reorthogonalizing the columns of the block vector $X_e$.
With the modes themselves in hand one can then construct
approximate propagators and ask how well these small eigenvalue
modes reproduce the full answers we obtain by conventional means.

\begin{figure}[hbtp]
\centerline{\psfig{figure=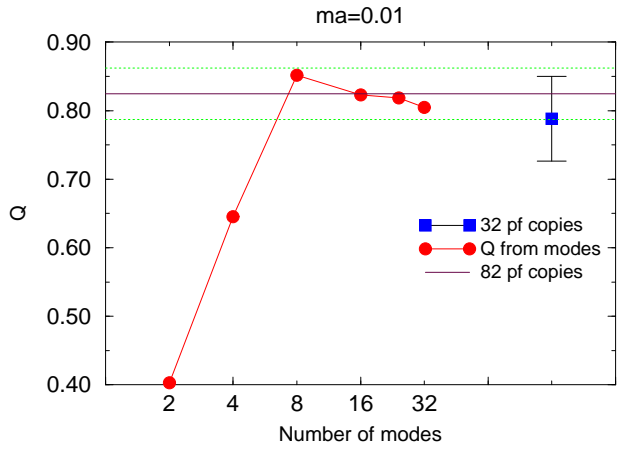,width=9.0cm}}
\caption{Topological charge on a typical dynamical configuration}
\label{q05}
\end{figure}

In figure \ref{q05} we show the contribution to the topological
charge coming from the eigenmodes in one typical dynamical configuration.
In terms of the eigenmodes, equation~\ref{qfrmu1} takes the form
\begin{equation}
Q = m\sum_{\lambda}{\frac{{\psi}_{\lambda}^{\dag}\,{\gamma}_5\,{\psi}_{\lambda}}{i{\lambda} + m}}.
\label{trg5sf1}
\end{equation}
As can be seen, after the first few modes, the sum quickly
saturates to the full answer obtained by averaging over
many copies of pseudofermion noise.

\begin{figure}[hbtp]
\centerline{\psfig{figure=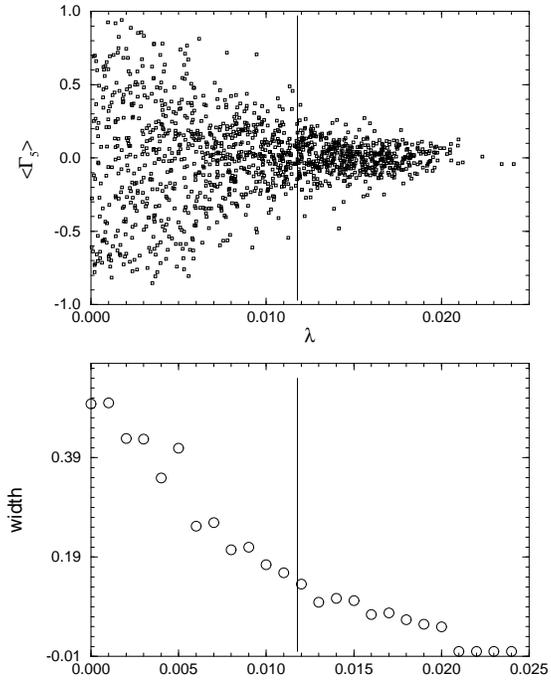,width=9.0cm}}
\caption{Above:$\langle {\Gamma}_5 \rangle$ on the dynamical ensemble. Below: Fluctuations in $\langle {\Gamma}_5 \rangle$.}
\label{fig51}
\end{figure}

That this behavior is typical of the whole ensemble can
be seen in the scatter plot of figure \ref{fig51}, which
plots the eigenvalue, $\langle\gamma_5\rangle$ pairs
collected on all configurations.  The largest contributions
to the trace evidently come from the smallest eigenmodes.
We conclude that even at this finite lattice spacing,
the index theorem is perfectly recognizable.

\begin{figure}[hbtp]
\centerline{\psfig{figure=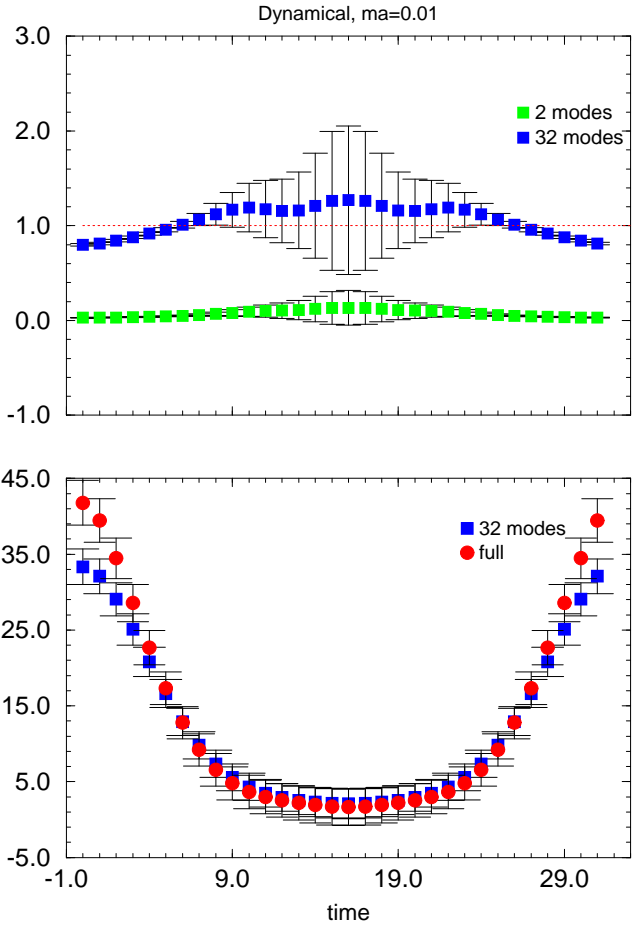,width=9.0cm}}
\caption{Below: Comparison of the two loop amplitudes. Above: Ratio of the full two-loop amplitude to that calculated with modes.}
\label{fig511}
\end{figure}

It is only natural to extend the analysis further to calculate the $\eta'$ 
disconnected correlator, which is proportional to the fluctuations
in the topological charge and is given by 
$\langle {\rm tr}({\gamma}_5\,G(x,x)){\rm tr}({\gamma}_5\,G(y,y))\rangle $.
The result of calculating this using the available eigenmodes is
shown in figure \ref{fig511}.  It is evident that the first few modes
give essentially the full answer. 

\begin{figure}[hbtp]
\centerline{\psfig{figure=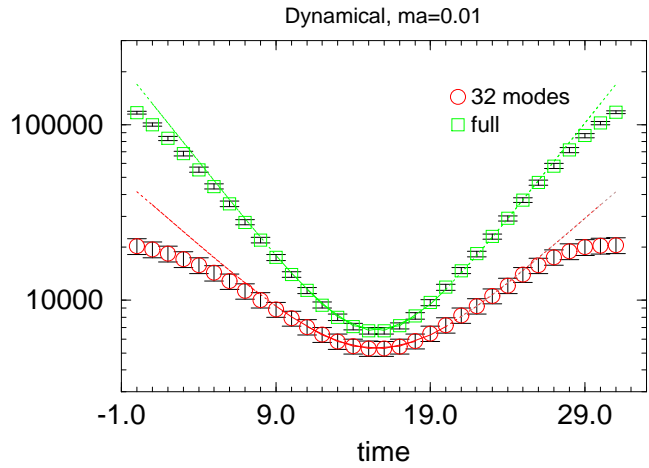,width=9.0cm}}
\caption{Pion propagator on dynamical ensemble at $ma=0.01$.}
\label{figpion}
\end{figure}

One may wonder if the chiral modes may be used to successfully 
reproduce other hadronic correlators as well.  In terms of the 
eigenmodes, the approximate propagator is of the form
\begin{equation}
G_{xy} \approx \sum_{\lambda} {\psi_\lambda(x) \psi^{\dag}_\lambda(y)
\over i\lambda +m}
\end{equation}
Figure \ref{figpion} answers this question for the case
of the Goldstone pion.  Even with the full 32 modes, the
pion propagator is off by a large factor both in its
overall normalization, and in its mass parameter.
Evidently the $\eta'$ is special in that it is exquisitely
sensitive to the lowest handful of modes.

\section{CONCLUSIONS}

We have computed both the quenched flavor singlet
vertex $m_0$ and the full dynamical $\eta'$ mass,
finding agreement with the experimental numbers.
Since the statistical errors are relatively large
we have not attempted to pin down the systematic
errors, e.g.  from finite lattice spacing and from
the neglect of $\eta$-$\eta'$ mixing.
Our results both confirm our conventional
understanding of the $\eta'$ in QCD and show
that lattice QCD can be a useful tool in this
difficult sector.  We have also shown that unlike
other hadron correlators, the $\eta'$ propagator
is particularly sensitive to the presence of 
fermionic zeromodes.  This result supports the
connection of the $\eta'$ to topology.

\begin{table}[tbhp]
\caption{The Statistical Ensemble}
\hspace{5pt}
\label{TAB1}
\begin{tabular}{|c|c|c|c|c|}
\hline
$N_f$  & $m_{dyn}$ & $\beta$  & $N_{samp}$ & $m_{val}$ \\
\hline
0      &  $\infty$ & 6.0      &  83          &  0.011     \\
       &           &          &              &  0.022     \\
       &           &          &              &  0.033     \\
2      &  0.01     & 5.7      &  79          &  0.01      \\
       &           &          &              &  0.02      \\
       &           &          &              &  0.03      \\
2      & 0.015     & 5.7      &  50          &  0.01      \\
       &           &          &              &  0.015     \\
2      & 0.025     & 5.7      &  34          &  0.01     \\
       &           &          &              &  0.025    \\
\hline
\end{tabular}  
\end{table}

\begin{table}[hbtp]
\begin{footnotesize}
\caption{Smeared valence operators. The spin/flavor is ${\gamma}_5 \otimes I$
in each case.}
\label{TAB2}
\begin{center}
\begin{tabular}{|l|c@{\hspace{6pt}}c|c|c|} \hline
   &            Contraction&             & ${\kappa}^2$ at source             & ${\kappa}^2$ at sink \\ \hline
LL & $\overline{Q}(x)Q(x)$&$\overline{Q}(0)Q(0)$          & no smearing & no smearing \\
$C_1$ & $\overline{Q}(x)Q(x)$&$\overline{Q}_{w4}(0)Q(0)$     & -0.6 & no smearing \\
$C_2$ & $\overline{Q}_{w1}(x)Q(x)$&$\overline{Q}_{w4}(0)Q(0)$&  -0.6 & -0.5 \\
$C_3$ & $\overline{Q}_{w2}(x)Q(x)$&$\overline{Q}_{w4}(0)Q(0)$& -0.6 & -0.53 \\
$C_4$ & $\overline{Q}_{w3}(x)Q(x)$&$\overline{Q}_{w4}(0)Q(0)$&  -0.6 & -0.56 \\
$C_5$ & $\overline{Q}_{w4}(x)Q(x)$&$\overline{Q}_{w4}(0)Q(0)$&  -0.6 & -0.6 \\
\hline
\end{tabular}
\end{center}
\end{footnotesize}
\end{table}

\begin{table}[hbtp]
\caption{$m_{\eta'}(N_f=3)$ in the chiral limit.}
\label{TAB3}
\begin{tabular}{|ccc|}
\hline
        &Quenched     &Dynamical ($m_{dyn}=m_{val}$)\\
        &(MeV)        &(MeV)   \\
\hline
LL & 1156(95)   & 974(133) \\
${\rm LLC_5}$ & 891(101)   & 780(187) \\
\hline
\end{tabular}
\vspace{-0.4cm}
\end{table}

\begin{table}[hbtp]
\caption{$\alpha$ from the quenched ratio data.}
\label{TAB4}
\begin{tabular}{|ccc|}
\hline
$m_{val}\,a$ & LL  & $LLC_5$ \\
\hline
0.011 & -1.7496(2321) & -0.2988(2091) \\
0.022 & -0.8475(1301) & -0.1431(1195) \\
0.033 & -0.4645(910) & -0.0399(702) \\
\hline
\end{tabular}
\vspace{-0.1cm}
\end{table}

\end{document}